\documentclass{PoS}

\pdfoutput=1
\usepackage{natbib}
\usepackage{xspace}
\usepackage{graphicx}

\bibpunct[; ]{[}{]}{,}{a}{}{;}

\newcommand{\askap}{{\sc Askap}\xspace}

\newcommand{\dingo}{{\sc Dingo}\xspace}
\newcommand{\wallaby}{{\sc Wallaby}\xspace}
\newcommand{\gama}{{\sc Gama}\xspace}

\newcommand{\herschelsurvey}{Herschel-{\sc Atlas}\xspace}

\newcommand{\galex}{{\sc Galex}\xspace}

\newcommand{\vst}{{\sc Vst}\xspace}

\newcommand{\ukirt}{{\sc Ukirt}\xspace}

\newcommand{\vista}{{\sc Vista}\xspace}

\newcommand{\wise}{{\sc Wise}\xspace}

\newcommand{\aaomega}{{AA$\Omega$}\xspace}
\newcommand{\hi}{{\sc Hi}\xspace}
\newcommand{\ohi}{$\Omega_{\rm HI}$\xspace}

\newcommand{\hipass}{{\sc Hipass}\xspace}
\newcommand{\ska}{{\sc Ska}\xspace}

\title{Exploring the HI Universe with ASKAP}

\ShortTitle{DINGO}

\author{\speaker{Martin Meyer}\\
        International Centre for Radio Astronomy Research\\
        M468, The University of Western Australia\\
        35 Stirling Highway, Crawley 6009, Australia\\
        E-mail: \email{martin.meyer@icrar.org}}

\author{The DINGO team\\
       http://www.physics.uwa.edu.au/$\sim$mmeyer/dingo}

\abstract{The survey speed of \askap makes it a prime instrument with which to survey the \hi universe, enabling it to carry out both wide surveys of the entire sky, as well as deep surveys covering cosmologically representative volumes.  Here, the use of \askap to study deep \hi fields is discussed as proposed by the Deep Investigation of Neutral Gas Origins (\dingo) survey.  This \askap science survey project anticipates observing in excess of $10^5$ sources out to redshift $z\sim0.4$.   Key science goals include: $\Omega_{\rm HI}$ and its evolution, the cosmic web as traced by distributions such as the \hi mass function and the 2pt correlation function, and the formation and evolution of galaxies.  Science returns are maximised by targeting the \gama survey regions, enabling the \hi content of galaxies to be studied and understood in full context with all the major galactic constituents over the past 4 Gyr.}

\FullConference{Panoramic Radio Astronomy: Wide-field 1-2 GHz research on galaxy evolution - PRA2009\\
		 June 02 - 05 2009\\
		 Groningen, the Netherlands}

\begin{document}

\section{Introduction}

\noindent Understanding how galaxies form and evolve is one of the key astrophysical problems of the 21$^{\rm st}$ century.  Recent observations of galaxies, supernovae, and the Cosmic Microwave Background have refined measurements of key cosmological parameters and simulations have accurately modelled how dark matter halos collapse from the miniscule density perturbations carried over from the inflationary era. But understanding how baryons collapse into these potential wells and ultimately form stars, how black holes are formed, how the baryons interact with the dark matter, what the balance is between accretion and outflow, heating and cooling, are problems which require a deep understanding of physics on an immense range of spatial scales and particle densities. These problems remain largely unsolved.

One of the keys to a deeper understanding is to observe and model the gaseous component of the Universe.  Baryons are believed to have existed entirely in gaseous form before the epoch of galaxy formation, and it is by the continual collapse of this gas into the filamentary structures of the cosmic web, its accretion into the deepest potential wells, the creation of cool gaseous disks and their ultimate collapse into dense molecular clouds, that galaxies are able to form stars.  The on-going feedback between gas and stars, and the governing role of the host dark matter potentials, continue to determine the evolution of galaxies today.  To date, our ability to observe the gaseous content of the universe, and hence to explain the manner in which these processes have occurred, is extremely limited.  Observations have largely been restricted to small samples of gas in emission in the local universe, or to sparse samples of absorption line systems at higher redshifts, and the high level of uncertainty inherent in interpreting such observations.  

\section{\dingo and \askap}

\noindent The Australian \ska Pathfinder telescope (\askap) \cite{Feain2010} has been designed as a survey instrument and is particularly well-suited to the search for 21cm neutral hydrogen emission.  Its large field of view (30 deg$^2$), in combination with its collecting area (4072 m$^2$) and system temperature (50K) give it a fast survey speed, enabling the telescope to go both wide and deep in a comparatively short period of time.  Its wide 300 MHz instantaneous bandwidth similarly facilitates surveys of cosmologically representative volumes in \hi.  

Where \wallaby will perform a shallow survey of the southern sky, \dingo aims to provide a legacy deep dataset for \hi emission out to $z\sim0.4$.   To enable multiwavelength studies, \dingo will target the \gama survey areas \cite{Driver:2010p8129}. \gama is composed of several imaging and spectroscopic surveys using the world's pre-eminent facilities, spanning all wavelengths from the ultraviolet to the far infrared (\galex, \vst, \vista, \ukirt, \wise, \herschelsurvey), in addition to 250k \aaomega spectra in key overlap regions.  A tiered \hi survey strategy is proposed covering regions 150 deg$^2$ (500 hours per pointing, $0.15<z<0.29$) and 60 deg$^2$ (2500 hours per pointing, $0.1<z<0.43$) to meet the science goals described in the following section.  

\dingo is a major \ska pathfinder experiment and will provide an ideal dataset for \ska planning.  This will be acheived scientifically through its study of the evolving \hi universe, and technologically through the use of \askap's phased-array feeds for extended integrations.

\begin{figure}[h]
\footnotesize
\begin{center}
\includegraphics[trim=0mm 0mm 0mm 0mm,width=15.5cm,keepaspectratio=true,clip]{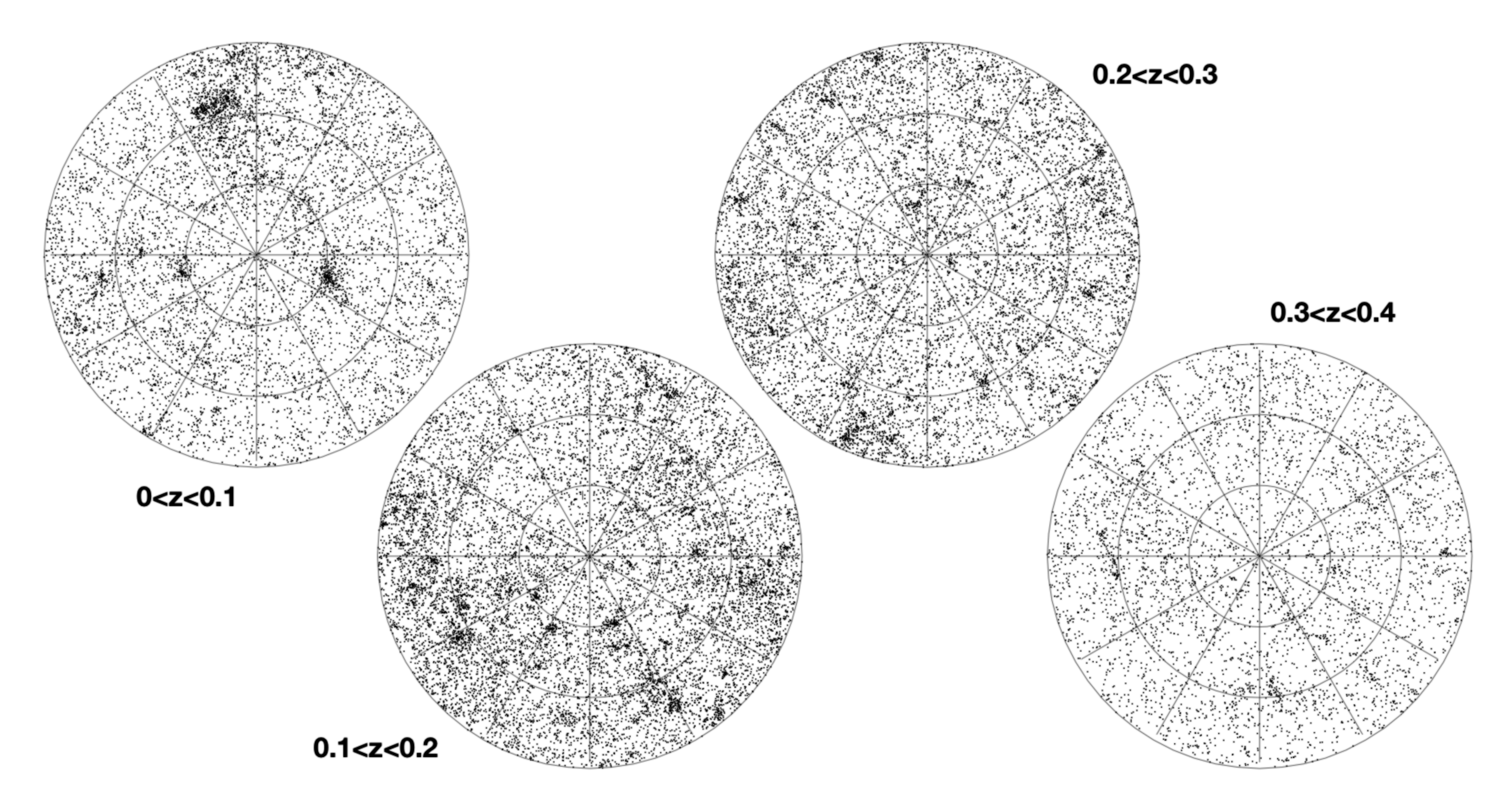}
\caption{Simulated spatial distribution of galaxies in a 2500 hr, 30 deg$^2$ \askap pointing.  Our proposed ultradeep low-z data includes two fields $0.1<z<0.43$ (2500 hr integrations) while the deep includes five fields $0.15<z<0.29$ (500 hr integrations) to provide adequate cosmic volumes and galaxy numbers for dependency analysis.}
\label{fig:spatial}
\end{center}
\vspace{-5mm}
\end{figure}

\section{Key Science Goals}

\noindent The core \dingo science themes and goals are:
\begin{itemize}
\item Evolution of the cosmic \hi density: On the largest scales, evolution of the gaseous universe is traced by measuring changes in the cosmic gas-mass density with time.  The \hi Parkes All-Sky Survey (\hipass) \cite{Meyer:2004p62} and other large blind \hi surveys have provided accurate measurements of the local \hi cosmic mass density \cite{Zwaan:2005p65}. However, beyond $z=0$ the sheer difficulty of \hi detection has limited our ability to detect \hi in emission in individual galaxies, necessitating the use of stacking techniques \cite{Lah:2007} and a hetergeneous mix of methods at the highest redshifts \cite{Rao:2006p4834,Prochaska:2009p5679} with significant uncertainty in both measurement and interpretation.  \dingo aims to measure \ohi directly through \hi emission with \hipass-like errors out to $z\sim0.4$
\item The cosmic web: As well as changes in the \hi cosmic mass density, it is also important to understand the way in which the distribution of \hi gas has changed with time.  This has varied significantly over the history of the universe, evolving from a near uniform distribution with only small density fluctuations at the start of the dark ages, to its current highly clustered state, mostly restricted to galactic halos where its density is high enough to self-shield.  \dingo will examine the most recent changes in the \hi mass distribution through evolutionary measurements of the \hi mass function \cite{Zwaan:2005p65}, the 2pt correlation function \cite{Meyer:2007p64}, and HOD modelling \cite{Wyithe2009}.  The simulated spatial distribution of galaxies in a single \dingo ultradeep field is shown in Figure ~\ref{fig:spatial}.
\item Galaxy evolution: Large, comprehensive datasets with galaxy-by-galaxy data will ultimately be required to fully understand the baryonic processes in galaxies.  \gama is the best multi-wavelength dataset in existence over the areas and depths relevant to \dingo, and by targeting these survey regions, \dingo will enable an understanding of the evolution of \hi in the context of all the other major galactic constituents over the past 4 Gyr.  Our models based on current \askap specifications indicate that \dingo will be sensitive to ${\rm M}_{\rm HI}^*$ galaxies over the entire redshift range probed.  
\end{itemize}

\section{Design Study}

\noindent \dingo is now in the design study phase of the \askap survey science process.  During this time, a number of scientific and technical aspects necessary to making \dingo a success will be examined.

\subsubsection*{Survey Design }

\noindent A core task of the design study will be to determine the location of the \dingo fields within the \gama survey regions.  We will additionally examine other survey parameters during the design study to ensure that the final \dingo configuration will best meet the aims outlined in the science case using simulations, observations and analytic methods.  Additional studies will be made of key issues such as cosmic variance and the survey volumes required for environmental analysis.

\subsubsection*{Simulated Catalogues and Mock Skies}

\noindent Simulated \hi datasets provide a central means for evaluating the likely science outcomes for \dingo. To date, our simulations have been a combination of non-evolving semi-empirical methods taking into account the locally measured cosmic mass density and the \hi mass function, along with basic semi-analytic methods for the spatial distribution of galaxies.  During the design study, we plan to advance on these through the use of fully semi-analytic methods such as the recent work of Obreschkow et al. \cite{Obreschkow:2009p5486}.   Given the current relatively high mass limit of semi-analytic models, it may remain necessary to combine semi-analytic and semi-empirical approaches to deliver simulations covering the full mass spectrum, although we will seek to advance work on this issue as far as possible.  These catalogues will be developed in parallel to those proposed for \wallaby and we will investigate a number of simulation issues:

\begin{itemize}
\item an analysis of semi-empirical and semi-analytic methods to yield simulations consistent with known observational constraints on the cosmic \hi density, the \hi mass function, the clustering properties of \hi rich galaxies, and correlations between \hi properties and environment
\item simulations that cover a range of plausible evolutionary scenarios 
\item mock science analyses based on these simulated catalogues to examine our sensitivity, anticipated errors, and discriminatory power for separating various evolutionary models
\item assistance in the development of mock skies for \askap pipeline testing
\end{itemize}

\subsubsection*{Radio Frequnecy Interference and Cube Combination}

\noindent The deep observations of \dingo make RFI an issue that will require special attention.  The stacking of multiple observations will mean that RFI features may only become apparent some time after observations have begun.  A methodology needs to be developed to deal with this given the limited current storage capability of visibility data. 

\subsubsection*{Source Finding and Parametrization}

\noindent Multiple approaches to source detection need to be examined and tested. It is likely that smoothing with a matched filter or wavelet filtering may improve the detectability of point-like sources, which will be the vast majority of our likely detections. The processing overheads for these approaches will need to be evaluated, as the wavelet filtering in particular can be computationally intensive.  One of the primary tasks will be to test the completeness and reliability of the detections made with the source-finder. Due to the automated nature of the processing, the source-finder needs to be as complete (ie. maximise the number of real sources found) and reliable (ie. minimise the number of non-real sources identified) as possible, and we will develop tools to measure both these quantities.  \dingo will make use of, and contribute to, the standard end-to-end simulation/imaging/analysis pipeline under development by the \askap computing group, which will enable us to incorporate different simulations and observational strategies into our testing, and to provide uniform test results to aid the analysis. 

\subsubsection*{Early Science Commissioning}

\noindent Critical to early \askap commissioning are tests of: receiver performance and stability; beam performance and stability; calibration accuracy; techniques to ensure spatial and spectral invariance of calibration, including across beams; quality of sky models; dynamic range (continuum and spectral line); spectral bandpass stability (as a function of time, beam, position and polarization angle); mosaicing strategies for full beam combination; measurement of sidelobe response, including at equatorial latitudes; and the characterization of map noise as a function of angle and integration time.  This last issue is particularly critical given the deep nature of this survey.  BETA will be used as a valuable testbed for assessing these issues.

\subsubsection*{Ancillary Datasets}

\noindent Multi-wavelength data is vital for meeting the \dingo science goals.  We will monitor the development of the \gama datasets to ensure that the final \dingo field locations are best positioned to take advantage of the data products that are ultimately available, and to fill any gaps that arise where feasible and necessary.  

\subsubsection*{SKA Pathfinder Coordination}

\noindent There are a large number of major radio astronomy facilities currently undergoing simultaneous development, including APERTIF, the eVLA, ATA, MeerKAT, and of course \askap itself.  The \askap legacy surveys will not be operating in isolation, and numerous major programs can be expected on the other pathfinder facilities.  In developing an \hi survey strategy, it is important to best target legacy surveys in ways that exploit the advantages of each facility and ideally operate in a complimentary rather than overlapping manner.  MeerKAT will also be an ideal instrument with which to survey distant \hi, and in combination with \askap these facilities have the opportunity to develop a layered survey strategy that will be able to provide a comprehensive view of the universe, starting with \wallaby over the largest areas $\sim 3\times10^4$ deg$^2$, \dingo over areas 60-150 deg$^2$, and MeerKAT surveying areas $\sim$ 10 deg$^2$ and less.  The exact parameters and configurations of these future facilities continue to evolve, as do the projects their user communities seek to carry out, and as such it is important to maintain collaborative links to ensure the overall science outcomes are maximised.   

\bibliographystyle{JHEP}
\bibliography{dingo}



\end{document}